\documentstyle{ichep}
\newcommand{\be}{\begin{equation}}
\newcommand{\ee}{\end{equation}}
\newcommand{\bea}{\begin{eqnarray}}
\newcommand{\eea}{\end{eqnarray}}
\newcommand{\nn}{\nonumber}

\begin{document}

\title{$B \rightarrow X_s\,\gamma$ and $B \rightarrow K^*\,\gamma$
in the standard and 2H models}

\author{Marco Ciuchini$^{\dag}$}

\affil{INFN, Sezione Sanit\`a, \\
V.le Regina Elena 299, 00161 Roma, Italy}

\abstract{Theoretical predictions for the branching ratios of the
$B \rightarrow X_s\,\gamma$ and $B \rightarrow K^*\,\gamma$ decays are
calculated in the Standard Model and in the (type II) two-Higgs-doublet model.
Both the complete leading and the partially known next-to-leading order QCD
corrections are included. The uncertainties due to the regularization scheme
dependence introduced by the incomplete NLO terms are discussed. The results
are compared with the recent CLEO II measurements and a new lower limit on the
charged Higgs boson mass, $M_{H^\pm} > \sim 200$ GeV, is obtained.}

\twocolumn[\maketitle]

\fnm{1}{E-mail: ciuchini@vaxsan.iss.infn.it.}

We calculate the theoretical predictions of the branching ratios for the
decay $B\to X_s\gamma$ and $B\to K^*\gamma$ in the Standard Model and
the type II two-Higgs-doublet model \cite{charged}. It is well known
that the leading order (LO) QCD corrections are important \cite{gsw},
almost doubling the amplitudes of these decays.
We include QCD correction using the results of ref. \cite{cfmrs1} to compute
the
relevant Wilson coefficient, $C_7^{eff}(\mu)$.
We also include those next-to-leading (NLO) corrections which are already known
\cite{bjlw,cfmr2}, as explained in ref. \cite{cfmrs2}.
This results in a significant reduction of the dependence of
$C_7^{eff}(\mu)$ on the renormalization scale $\mu$, which is
the main source of theoretical uncertainty in the leading order calculation
\cite{bmmp}, see \fref{fig:c7}.
However this procedure is not consistent theoretically and, in fact, an
unphysical
regularization scheme dependence is introduced in the physical predictions in
this
way. We try to cope with this problem by considering two different cases, the
$\overline{{\rm MS}}$ 't Hooft-Veltman (HV) and naive dimensional (NDR)
regularization/renormalization
schemes, taking, for each prediction, the mean value over the two schemes as
the
physical result. Moreover, the difference between them is assumed as a
systematic
error associated to our ignorance of the full next-to-leading corrections.
This error is presented along with the usual one, due to the variation of the
relevant parameters, $\Lambda_{QCD}$ and $m_t$.

\Figure{14pc}{LO and NLO $C_7^{eff}$ as a function of $\mu$.\label{fig:c7}}

\begin{table}
\Table{|c|c|}{
\hline
Parameter & Value \\ \hline
$\vert V_{ts}^* V_{tb} \vert^2/\vert V_{cb}\vert^2$ & $0.95 \pm 0.04$ \\
\hline
$m_c/m_b$ & $0.316\pm 0.013$ \\ \hline
$m_t$ (GeV) & $174 \pm 17$ \\ \hline
$\lambda_1$ (GeV$^2$) & $-0.15\pm 0.15$ \\ \hline
$\lambda_2$ (GeV$^2$) & $0.12 \pm 0.01$ \\ \hline
$m_b(\mu=m_b)$ (GeV) & $4.65 \pm 0.15$ \\ \hline
$F_1(0)$ & $ 0.35 \pm 0.05$ \\ \hline
$BR(B \rightarrow X l \nu_l)$ & $0.107 \pm 0.005$ \\ \hline
$\Lambda_{QCD}^{n_f=4}$ (MeV) & $330 \pm 100$ \\ \hline
$\mu$ & $m_b/2$--$2 m_b$ \\ \hline}
\caption{Values of the parameters used to predict
the radiative B decay rates.}
\label{tab:par}
\end{table}

The relevant formulae to calculate the branching ratios we are interested in
are\\
\\
{\small
$BR(B \rightarrow X_s \gamma )= \left[ \frac{\Gamma(
B \rightarrow X_s \gamma)}{\Gamma(B \rightarrow X l \nu_l)} \right]
BR(B \rightarrow X l \nu_l)$, \\
$\phantom{mm}\left[ \frac{\Gamma(
B \rightarrow X_s \gamma)}{\Gamma(B \rightarrow X l \nu_l)} \right]=
\frac{\vert V_{ts}^* V_{tb} \vert^2}{\vert V_{cb}\vert^2} \frac{\alpha_e}
{6 \pi g(m_c/m_b)} F \vert C_7^{eff}(\mu) \vert ^2$,\\
$\phantom{mm}g(z)=1-8 z^2 +8 z^6-z^8-24 z^4 \ln(z)$,
$F=\frac{K(m_t/M_W,\mu)}{\Omega(m_c/m_b,\mu)}$.
\\
\\
$BR(B \rightarrow K^* \gamma )=
\left[ \frac{\Gamma(B \rightarrow K^* \gamma)}
{\Gamma(B \rightarrow X_s\gamma)} \right]
\left[ \frac{\Gamma(B \rightarrow X_s \gamma)}
{\Gamma(B \rightarrow X l \nu_l)} \right]
BR(B \rightarrow X l \nu_l)$\\
$\phantom{mm}\left[ \frac{\Gamma(B \rightarrow K^* \gamma)}
{\Gamma(B \rightarrow X_s \gamma)} \right]=
\left(\frac{M_b}{m_b}\right)^3\left(1-\frac{M_{K^*}^2}{M_B^2}\right)^3
\frac{\vert F_1(0) \vert^2}{1 +
(\lambda_1-9 \lambda_2)/(2 m_b^2)}$
}
\\

\noindent $BR(B \rightarrow X_s\gamma )$ includes also the known
next-to-leading
corrections to the matrix element, while non-perturbative $1/m_b^2$
corrections are included in $BR(B \rightarrow K^* \gamma )$.
The numerical values of the different quantities appearing in these expressions
are given in \tref{tab:par}. For more details on their choice, see ref.
\cite{cfmrs2}.

\begin{table}[b]
\Table{|c|c|c|c|}{
\hline
$\phantom{\mu}$ & \multicolumn{3}{|c|}{$BR(B \rightarrow X_s \gamma)\times
10^4$} \\ \hline
$\mu$ (GeV) & LO & NLO$_{HV}$ & NLO$_{NDR}$ \\ \hline
$m_b/2$ & $3.81\pm 0.47$ & $1.92\pm 0.19$ & $2.77\pm 0.32$ \\ \hline
$m_b$ & $2.93\pm 0.33$ & $ 1.71\pm 0.18$ & $2.25\pm 0.25$ \\ \hline
$2 m_b$ & $2.30\pm 0.26$ & $ 1.56\pm 0.17$ & $1.91\pm 0.21$ \\ \hline \hline
$\phantom{\mu}$ & \multicolumn{3}{|c|}{$BR(B \rightarrow K^* \gamma)\times
10^5$} \\ \hline
$\mu$(GeV) & LO & NLO$_{HV}$ & NLO$_{NDR}$ \\ \hline
$m_b/2$ & $6.9\pm 1.5$ & $4.4\pm 0.8$ & $6.4\pm 1.3$ \\ \hline
$m_b$ & $5.3\pm 1.1$ & $3.8\pm 0.8$ & $5.0\pm 1.0$ \\ \hline
$2 m_b$ & $4.2\pm 0.9$ & $ 3.3\pm 0.7$ & $4.1\pm 0.8$ \\
\hline}
\caption{Theoretical predictions of the radiative branching ratios.}
\label{tab:br}
\end{table}

Using the previous formulae, we calculate the branching ratios in
\tref{tab:br}. The errors shown in this table are due to the uncertainties on
$\Lambda_{QCD}$ and $m_t$. Combining the NLO results in HV and NDR for
different
values of $\mu$, we obtain our final predictions in the Standard Model
\bea
BR(B \rightarrow K^* \gamma) &=& (4.3\pm 0.9^{ +1.4}_{-1.0}) \times 10^{-5}
\nn\\
BR(B \rightarrow X_s \gamma) &=& (1.9 \pm 0.2\pm 0.5) \times 10^{-4} \nn \\
\frac{\Gamma(B \rightarrow K^* \gamma)}{\Gamma(B \rightarrow X_s \gamma)}
&=& 0.23 \pm 0.09, \nn
\eea
Comparing them with the recent measurements \cite{cleoiie,cleoiii}
\bea
BR(B\rightarrow K^*\gamma)&=&(4.5\pm 1.5\pm 0.9)\times 10^{-5}\nn\\
BR(B\rightarrow X_s\gamma)&=&(2.32\pm 0.51\pm 0.29\pm 0.32)\times 10^{-4},\nn
\eea
a very good agreement is found. Notice, however, that the estimate of the
exclusive
branching ratio strongly depends on the value assumed for the form factor
$F_1(0)$.

Finally, let us consider the two-Higgs-doublet model known in the literature
as Model II \cite{charged}. Two more free parameters are present in this model,
$M_{H^\pm}$ and $\tan\beta$.
The charged Higgs boson exchange only modifies the initial conditions of the
Wilson coefficients. Moreover, for $\tan\beta > 1.5-2$, these become
practically independent of $\tan\beta$. In \fref{fig:2h}, the
$BR(B \rightarrow X_s \gamma)$ is reported as a function of $M_{H^\pm}$ for
$\tan \beta=2$. The band accounts for the theoretical uncertainties. The
comparison
with the experimental result gives a limit on $M_{H^\pm} >\sim 200$ GeV.

\Figure{14pc}{Predictions for  $BR(B \rightarrow X_s \gamma)$
in the 2H model with $\tan \beta=2$ are given as a function of
$M_{H^\pm}$. The experimental band is delimited by the dotted
lines.\label{fig:2h}}

\section*{Acknowledgments}
It is a pleasure to acknowledge the friendly collaboration of E. Franco,
G. Martinelli, L. Reina and L. Silvestrini.

\end{document}